**Electron trajectories and radiation growth rate in free electron laser with electromagnetic wiggler**

H. Ehsani[1], T. Mohsenpour[2]

[1] Department of Physics, Islamic azad University, nour branch, Mazandaran, Iran

[2] Department of Physics, Faculty of Basic Sciences, University of Mazandaran, Babolsar, Iran

E-mail: h_ehsani@iaunour.ac.ir

**ABSTRACT**

An analysis of steady-state electron trajectories by simultaneous solution of the equation of motion and the dispersion relation (DR) for electromagnetic wave wiggler in free-electron laser (FEL) with axial magnetic field is presented. The effects of the normalized axial magnetic field and the normalized angular frequency of electromagnetic wave wiggler on axial and transverse velocity for group I and II orbits are investigated. A fluid model is used to obtain the DR for electrostatic wave and the right and the left circularly polarized electromagnetic waves with all relativistic effects included. This dispersion relation is solved numerically to investigation the unstable coupling among all waves. When the transverse velocity is small, only the FEL instability is found. In group II orbits, with large transverse velocity, new coupling between the negative and positive energy space charge waves as well as between the left circular wave and positive energy space charge wave are found.

## 1. INTRODUCTION

The free electron laser (FEL) has been shown to be a promising source of radiation over a broad range of frequencies. The physical mechanism in FEL depends upon the propagation of an electron beam through a periodic magnetic field (wiggler). There are several types of wigglers called electrostatic, magnetostatic and electromagnetic wigglers. The wavelength of the output radiation for an electromagnetic wave wiggler (EWW) scales as $\lambda \approx \lambda_w / 4\gamma_0^2$ while for a magnetostatic wiggler it scales as $\lambda \approx \lambda_w / 2\gamma_0^2$ [1]. $\lambda_w$ is the wiggler wavelength (period) and $\gamma_0 = (1 - v_0^2/c^2)^{-1/2}$. As a result, in equal conditions, the EWW will generate radiation with shorter wavelengths. Therefore, for a given electron beam energy, EWW is advantageous over the magnetostatic wiggler for the production of short wavelengths. It needs to be mentioned that for a given wavelength of the radiation, EWW requires smaller electron beam energy compared to a magnetostatic wiggler [1-4]. Theoretical and numerical studies of a FEL with a magnetostatic wiggler have been studied by a number of researchers[5-14]. The FEL with electromagnetic wave wiggler was first analyzed by Sprangle, and Granatstein [15]. There are also many studies on the electromagnetic wave wiggler FEL[16-24] and longitudinal electrostatic wiggler FEL[25-31].

In many papers, electrons dynamic behavior and chaos has been investigated [32,33]. In Ref. 34 the gain equation describing the interaction between an electron and the radiation field derived in the low-gain-per-pass limit. A theoretical study of electron trajectories, harmonic generation, and gain in a FEL with a linearly polarized electromagnetic-wave wiggler by Mehdian et al. [35] investigated for axial injection of electron beam. Olumi et al.[24] investigated the propagation of electromagnetic traveling wave in a FEL with an electromagnetic wiggler using the relativistic fluid Maxwell formulation. In Ref. 36 an analysis of equilibrium orbits for electrons by a simultaneous solution of the equation of motion and the dispersion relation for electromagnetic wave wiggler in a FEL with ion-channel guiding has been presented. Theoretical studies of electron trajectories and dispersion relations in a combined ion electrostatic field as well as large-amplitude backward-propagating electromagnetic waves analyzed by Mehdian and Jafari[37]. In Ref. 38 the dispersion relation for EWW ion channel free electron laser in the absence of left circular wave was derived.

In this manuscript, we derive the electron equilibrium orbits in FEL with an EWW and an axial magnetic field and investigate the equilibrium trajectories. We will also derive a general DR for the interaction of all the waves, in which all possible wave modes can have unstable couplings with each other. Contrary to the usual investigations, relativistic terms to all orders of the wiggler amplitude are retained in the linearized equations. This DR is solved numerically to find the unstable coupling between electrostatic and electromagnetic waves. The layout of the present work is as follows: In Sec. II, equilibrium orbits for the electron beam are calculated. In Sec. III, the equilibrium orbits are studied. In Sec. IV, a DR for FEL with EWW is obtained. In Sec. V, a numerical analysis is carried out to investigate the unstable couplings between waves. In Sec. VI, concluding remarks are made.

## 2. BASIC ASSUMPTIONS

In order to understand the FEL interaction it is first necessary to understand the individual electron motion in the presence of EWW as governed by the Lorentz force equation

$$\frac{d\mathbf{P}}{dt} = -e\left[\mathbf{E}_w + \frac{1}{c}\mathbf{v}\times(\mathbf{B}_w + \mathbf{B}_0)\right], \tag{1}$$

and

$$\frac{d\gamma}{dt} = -\frac{e}{m_0 c^2}\mathbf{v}\cdot\mathbf{E}_w \tag{2}$$

where $\mathbf{E}_w$ and $\mathbf{B}_w$ are electric and magnetic fields of circularly polarized EWW, which are described by

$$\begin{cases}\mathbf{B}_w(z,t) = B_w[\hat{\mathbf{e}}_x \cos(k_w z + \omega_w t) + \hat{\mathbf{e}}_y \sin(k_w z + \omega_w t)], \\ \mathbf{E}_w(z,t) = -\frac{\omega_w}{ck_w}B_w[\hat{\mathbf{e}}_x \sin(k_w z + \omega_w t) - \hat{\mathbf{e}}_y \cos(k_w z + \omega_w t)],\end{cases} \tag{3}$$

$B_w$ denotes the amplitude of the EWW magnetic field and $(\omega_w, k_w)$ describe the frequency and wave number. For generality, a uniform axial magnetic field $\mathbf{B}_0 = B_0\hat{\mathbf{z}}$ is also assumed to be present. Note that in the limit of $\omega_w \to 0$ EWW reduces to that of the static, one dimensional magnetostatic wiggler. Steady-state solutions are obtained by requiring that $\gamma = \text{constant}$. In this manner, the electron velocity can be obtained as

$$\mathbf{v} = v_w[\hat{\mathbf{e}}_x \cos(k_w z + \omega_w t) + \hat{\mathbf{e}}_y \sin(k_w z + \omega_w t)] + v_\parallel \hat{\mathbf{e}}_z, \tag{4}$$

where $v_w$ and $v_\parallel$ are constants given by

$$v_w = \frac{\Omega_w(v_p + v_\parallel)}{\Omega_0 - (\omega_w + k_w v_\parallel)}, \tag{5}$$

$$\frac{1}{\gamma_0^2} = 1 - \frac{v_w^2}{c^2} - \frac{v_\parallel^2}{c^2}, \tag{6}$$

and $\Omega_{0,w} \equiv eB_{0,w}/\gamma_0 m_0 c$, and $v_p = \omega_w/k_w$.

## 3. EQUILIBRIUM AND STEADY-STATE ELECTRON TRAJECTORIES

To analyze the electron trajectories, we will use the Maxwell's equations for a cold beam with uniform density to obtain $\omega_w$ and $k_w$ satisfying a dispersion relation of the form

$$\omega_w^2 - c^2 k_w^2 + \frac{\omega_b^2 (\omega_w + k_w v_\|)}{\Omega_0 - \omega_w - k_w v_\|} = 0, \tag{7}$$

where $\omega_b = \sqrt{4\pi n_b e^2 / \gamma_0 m_0}$ is the beam plasma frequency ($n_b$ is beam density).

To show the dependency of the orbital velocities, requires the simultaneous solution of Eqs. (5), (6), and (7). Figure 1 shows the wave modes described by DR (7). Solid lines represent relevant wave modes for stable orbits with $v_\| > 0$ that are physically accepted solutions.

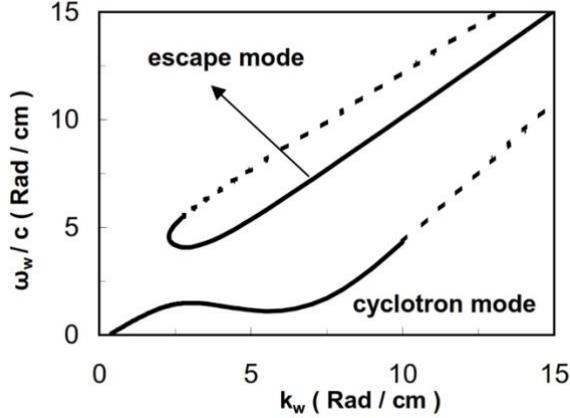

**Fig.1** Normalized angular frequency $\omega_w/c$ as a function of wave number $k_w$ for group I and group II orbits. Dotted line shows the unstable orbits

In this figure the parameters are $\Omega_0/c = 5.0263$, $\Omega_w/c = 0.1885$, $\gamma_0 = 2$, and $\omega_b/c = 1.3315$. The DR (7) contains two modes, i.e., escape mode and cyclotron mode. The solution corresponding to the escape mode is group I orbits with $\Omega_0 < k_w v_\| + \omega_w$ and the solution corresponding to the cyclotron mode is group II orbits with $\Omega_0 > k_w v_\| + \omega_w$. It should be noted that for escape mode, the phase velocity is larger than light velocity and for cyclotron mode, the phase velocity is smaller than light velocity. Figure 2 shows the variation of $v_\parallel/c$ as a function of the normalized cyclotron frequency $\Omega_0/k_w c$ for group I and group II orbits.

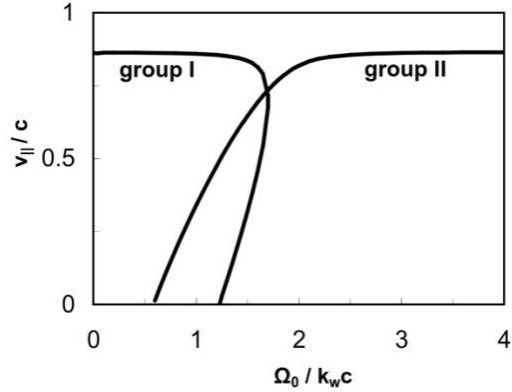

**Fig.2** Axial velocity $v_\parallel/c$ as a function of the normalized axial magnetic field $\Omega_0/k_w c$ for group I and group II orbits

In this figure the parameters are $\Omega_w/k_w c = 0.0839$, $\gamma_0 = 2$, and $\omega_b/k_w c = 0.212$. In Fig. 3, the relation between the normalized axial velocity $v_\parallel/c$ and the normalized angular frequency $\omega_w/c$ for both group I and group II orbits are shown. The range of the allowed values of $\omega_w$ for group II orbits are shown in Fig. 3.

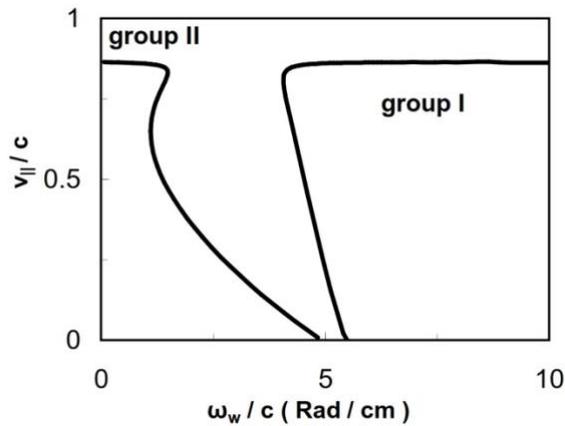

**Fig.3** Axial velocity $v_\parallel/c$ as a function of the normalized angular frequency $\omega_w/c$ for group I and group II orbits

The variation of $v_w/c$ with $\omega_w/c$ for group I and II orbits are shown in Fig. 4. In Figs. 3 and 4, the parameters are $\Omega_0/c = 5.0263$, $\Omega_w/c = 0.1885$, $\gamma_0 = 2$, and $\omega_b/c = 1.3315$.

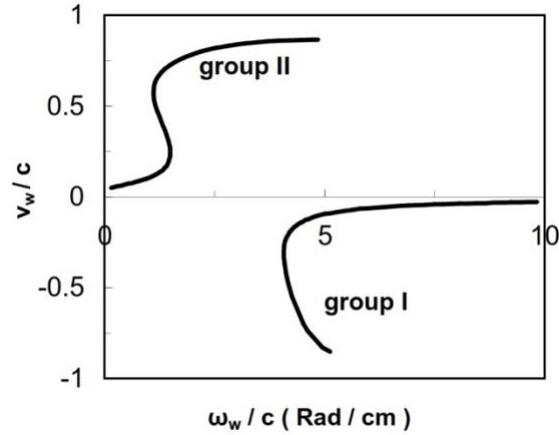

**Fig.4** Transverse velocity $v_w/c$ as a function of the normalized angular frequency $\omega_w/c$ for group I and group II orbits

Figure 4 shows that the value of $v_w$ increases with increasing the normalized cyclotron frequency for group I orbits while, it decreases for group II orbits. The variation of the wave number $k_w$ as a function of the normalized cyclotron frequency $\Omega_0/\omega_w$ for group I and group II orbits are shown in Fig. 5.

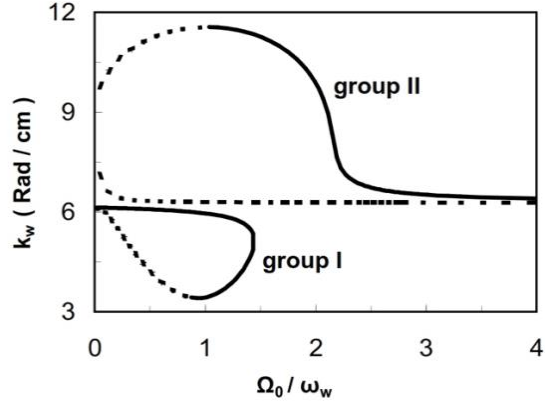

**Fig.5** Wave number $k_w$ as a function of the normalized axial magnetic field $\Omega_0/\omega_w$ for group I and group II orbits. Dotted line shows the unstable orbits. electron beam energy

This Figure shows that for group I orbits, the wave number is smaller in comparison with group II orbits. In Fig. 5, the parameters are $\Omega_w/c = 0.1885$, $\gamma_0 = 2$, and $\omega_b/c = 1.3315$. The variation of the wave number $k_w$ with the electron beam energy for group I and group II orbits is shown in Fig. 6.

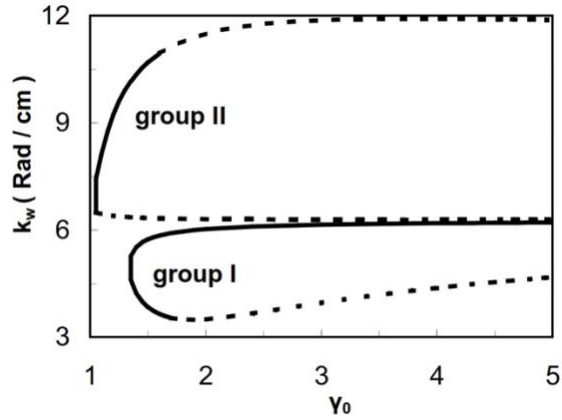

**Fig.6** Wave number $k_w$ as a function of the electron beam energy $\gamma_0$ for group I and group II orbits. Dotted line shows the unstable orbits

This figure shows that for $\omega_w = \text{constant}$, there is an upper limit on the electron beam energy for group I and II orbits. In Fig. 6 the parameters are $\Omega_0/c = 5.0263$, $\Omega_w/c = 0.1885$, and $\omega_b/c = 1.3315$.

## 4. DISPERSION RELATION

An analysis of the wave propagation in the FEL with an EWW and axial guiding magnetic field will be based on the electron continuity equation, the cold-electron relativistic momentum equation and the wave equation. The perturbed state can be considered as the unperturbed state $n_b = const.$, $\mathbf{v}_0 = \mathbf{v}_w + v_\parallel \hat{\mathbf{z}}$, $\mathbf{B}_w + \mathbf{B}_0$, and $\mathbf{E}_w$ plus small perturbations $\delta n$, $\delta \mathbf{v}$, $\delta \mathbf{E}$, and $\delta \mathbf{B}$. The unperturbed quantities are defined in the previous section. The perturbation quantities associated with the right and the left circularly polarized electromagnetic waves and the space charge wave are

$$\delta \mathbf{E}_{R\atop L}, \ \delta \mathbf{B}_{R\atop L}, \ \delta \mathbf{v}_{R\atop L} \sim (\hat{\mathbf{x}} \pm i\hat{\mathbf{y}}) \exp[i(k_{R\atop L} z - \omega t)], \tag{8}$$

$$\delta E_z, \delta v_z, \delta n_b \sim \exp[i(kz - \omega t)], \tag{9}$$

where $k_R = k - k_w$, $k_L = k + k_w$, $\omega_R = \omega + \omega_w$, $\omega_L = \omega - \omega_w$. By Substituting the perturbed quantities into the linearized continuity, relativistic momentum, and wave equations, a system of linear homogenous algebraic equations for the wave amplitudes, $\tilde{n}_b$, $\tilde{v}_z$, $\tilde{v}_R$, $\tilde{v}_L$, $\tilde{E}_z$, $\tilde{E}_R$, $\tilde{E}_L$, $\tilde{B}_R$, and $\tilde{B}_L$ are obtained. It is convenient to combine these equations to reduce the number of equations and unknowns versus components of the perturbed electric field. The necessary and sufficient condition for a nontrivial solution consists of the determinant of coefficients equated to zero. Imposing this condition yields

$$K_1(K_5 K_9 - K_6 K_8) + K_2(K_6 K_7 - K_4 K_9) + K_3(K_4 K_8 - K_5 K_7) = 0, \tag{10}$$

where

$$K_1 = A_R + \frac{A_R}{\omega_R}\{-k_R v_\parallel - \Omega_0 + A_1 + A_2\} + \omega_b^2\left\{1 - \frac{v_w^2}{2c^2} - \frac{k_R v_\parallel}{\omega_R}\right\}, \tag{11}$$

$$K_2 = \frac{A_L}{\omega_L}(-A_1 + A_2) - \omega_b^2 \frac{v_w^2}{2c^2}, \tag{12}$$

$$K_3 = \frac{kv_w}{\sqrt{2}}\{\omega_R - k_R v_\| - \Omega_0 + 2A_2\} + A_3(kv_\| - \omega) - \omega_b^2 \frac{v_w v_\|}{\sqrt{2}c^2}, \tag{13}$$

$$K_4 = \frac{A_R}{\omega_R}(A_1 - A_2) - \omega_b^2 \frac{v_w^2}{2c^2}, \tag{14}$$

$$K_5 = A_L + \frac{A_L}{\omega_L}\{-k_L v_\| + \Omega_0 - A_1 - A_2\} + \omega_b^2\left\{1 - \frac{v_w^2}{2c^2} - \frac{k_L v_\|}{\omega_L}\right\}, \tag{15}$$

$$K_6 = \frac{kv_w}{\sqrt{2}}\{\omega_L - k_L v_\| + \Omega_0 - 2A_2\} - A_3(kv_\| - \omega) - \omega_b^2 \frac{v_w v_\|}{\sqrt{2}c^2}, \tag{16}$$

$$K_7 = \frac{A_R}{\omega_R}\frac{\Omega_w}{\sqrt{2}}\left(1 + \frac{v_p v_\|}{c^2}\right) + \omega_b^2\left\{-\frac{v_\| v_w}{c^2 \sqrt{2}} + \frac{v_w}{\sqrt{2}}\frac{k_R}{\omega_R}\right\}, \tag{17}$$

$$K_8 = -\frac{A_L}{\omega_L}\frac{\Omega_w}{\sqrt{2}}\left(1 + \frac{v_p v_\|}{c^2}\right) + \omega_b^2\left\{-\frac{v_\| v_w}{c^2 \sqrt{2}} + \frac{v_w}{\sqrt{2}}\frac{k_L}{\omega_L}\right\}, \tag{18}$$

$$K_9 = -(\omega - kv_\|)^2 + \omega_b^2 \gamma_\|^{-2}, \tag{19}$$

$$A_1 = \frac{v_p v_w}{2c^2}\Omega_w, \tag{20}$$

$$A_2 = \frac{\gamma_\circ^2 v_w^2}{2c^2}k_w(v_p + v_\|), \tag{21}$$

$$A_3 = \frac{1}{\sqrt{2}}(k_w v_w + \Omega_w) + \frac{\gamma_\circ^2}{\sqrt{2}}\frac{v_w k_w v_\|}{c^2}(v_p + v_\|), \tag{22}$$

$$A_R = k_R^2 c^2 - \omega_R^2, \tag{23}$$

$$A_L = k_L^2 c^2 - \omega_L^2, \tag{24}$$

$$\gamma_\| = \left(1 - \frac{v_\|^2}{c^2}\right)^{-1/2}. \tag{25}$$

Equation (10) is the DR for the coupled electrostatic and electromagnetic waves propagating along a relativistic electron beam in the presence of an EWW and an axial guide magnetic field. As expected, the DR (10) reduces to the well-known results[12] of the magnetostatic wiggler in the limit as $\omega_w \to 0$. This DR (10) will be solved numerically, in the next section, to investigate the instabilities associated with different couplings between the wave modes.

The DR for the right wave alone, in the absence of the other two waves, is $K_1 = 0$ and the DR for the left wave alone, in the absence of the other two waves, is $K_5 = 0$ which indicate that the wiggler has direct effect on the right and left waves. Each DR consists of two modes, i.e., right-cyclotron ($R_c$) and right-escape ($R_e$) modes for the right wave; and left-cyclotron ($L_c$) and left-escape ($L_e$) modes for the left circularly polarized electromagnetic wave. $K_9 = 0$ is the DR for the space-charge wave when the right and left waves are absent. This shows that the space-charge wave is not directly affected by the wiggler. This because that the transverse motion of electrons, caused by the wiggler does not have any effect on the longitudinal oscillations of the space-charge wave. The DR for space-charge wave consists of two modes, i.e., positive-energy space-charge wave ($Sc_+$) with $\omega = kv_\parallel + \omega_b \gamma_\parallel^{-1}$, and negative-energy space-charge wave ($Sc_-$) with $\omega = kv_\parallel - \omega_b \gamma_\parallel^{-1}$.

## 5. NUMERICAL STUDIES

In order to find unstable couplings between the electrostatic and electromagnetic waves, the DR (10) will be solved numerically. Analyzing steady-state behavior on specific geometries with complex boundary conditions on electric or magnetic potential (Dirichlet) or their derivatives (Neumann) requires more advanced numerical methods such as the Finite Element Method, which has been applied in various fields of study to solve partial differential equations [39-48]. The parameters are $\Omega_w/k_w c = 0.0839$, $\gamma_0 = 3$, and $\omega_b = 0.212$. The roots of the DR (10) are found numerically for group I orbits. The normalized imaginary part of the coupling between the negative-energy space charge $(Sc_-)$ and the escape branch of the right circular wave $R_e$ as a function of the normalized $k/k_w$ for group I orbits with $\Omega_0/k_w c = 0.2$ is shown in Fig. 7.

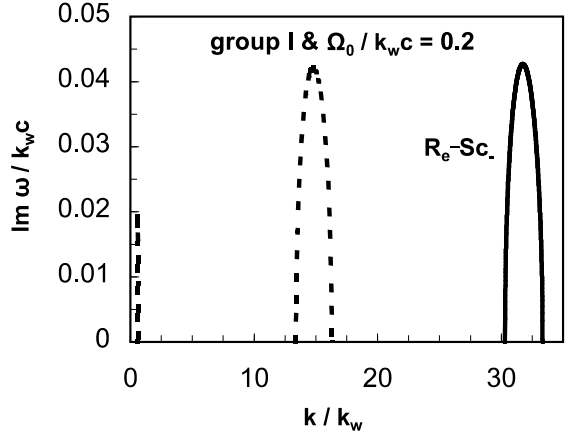

**Fig7** $\text{Im}\,\omega/k_w c$ versus $k/k_w$ for group I orbits. Dotted line shows the FEL instability in limit $\omega_w \to 0$

This coupling is called the FEL instability. In Fig. 7, the FEL instability in the limit $\omega_w \to 0$ ( the FEL with magnetostatic wiggler ) is shown with a dotted line for comparison. The FEL instability starts at $k/k_w \approx 30.3$ and ends at $k/k_w \approx 33.3$. The maximum growth rate of the FEL instability occurs at $k/k_w \approx 31.7$ with $\text{Im}(\omega/k_w c)_{max} = 0.043$ while, it occurs at $k/k_w \approx 14.8$ with $\text{Im}(\omega/k_w c)_{max} = 0.042$ for magnetostatic wiggler. Figure 7 shows that the FEL instability in EWW occurs at the shorter wavelength in comparison with the case of magnetostatic wiggler. The results of numerical solution of the DR (10) for group II orbits have been shown in Fig. 8 for $\Omega_0/k_w c = 4$, 3, and 2.65. It should be noted that, for group II orbits, with decreasing normalized cyclotron frequency, $\Omega_0/k_w c$ increases the transverse wiggler velocity, e. g., $v_w$ is 0.0773, 0.1434, and 0.1998 in Figs. 8 (a), 8(b), and 8(c), respectively.

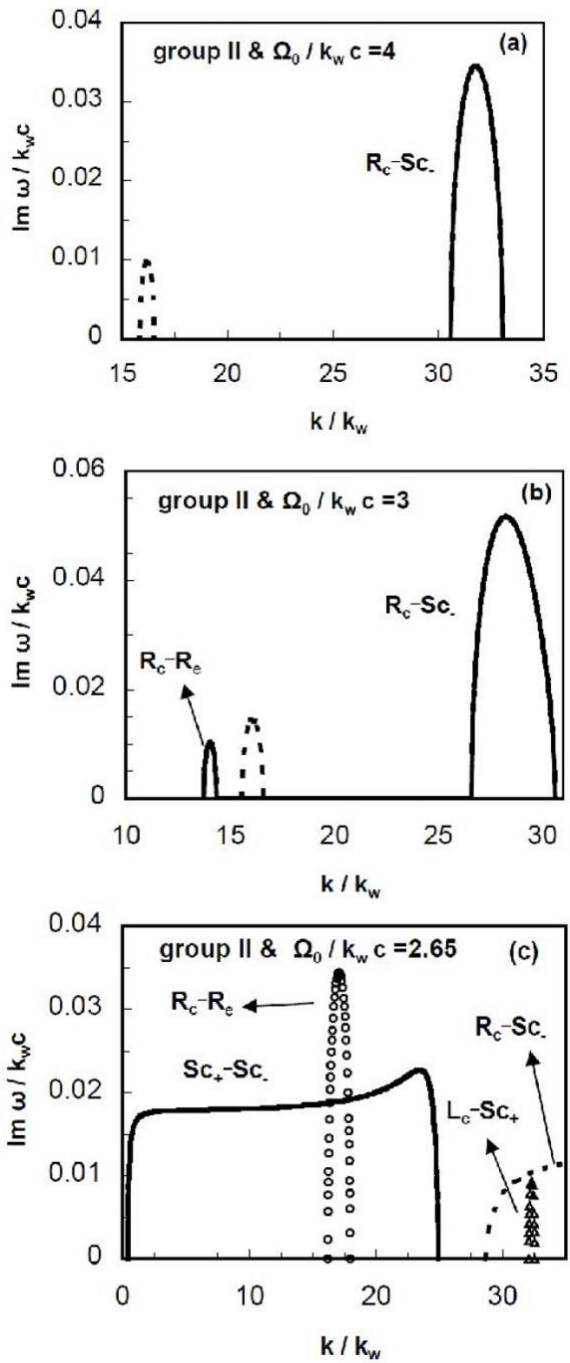

**Fig.8** $\mathrm{Im}\,\omega/k_w c$ versus $k/k_w$ for group II orbits. Dotted line in Figs. (a) and (b) show the FEL instability in limit $\omega_w \to 0$

$v_w$ plays important roles in the unstable coupling between waves. It is interesting to note that with increasing $v_w$, new unstable couplings are found, e.g., in Figs. 8(b) and 8(c). In Fig. 8(a), the normalized imaginary part of $\omega/k_w c$ versus $k/k_w$ for coupling between $R_c$ mode and $Sc_-$ mode is shown. This coupling starts from $k/k_w = 30.58$ and ends at $k/k_w = 33.06$ with $\text{Im}(\omega/k_w c)_{\max} = 0.0345$ at $k/k_w = 31.78$. In Fig. 8(a), the FEL instability in limit $\omega_w \to 0$ is shown with a dotted line for comparison. In Fig. 8(b), in addition to the FEL instability, a new coupling between the $R_c$ mode and the $R_e$ mode is found. Coupling between the $R_c$ and the $R_e$ modes starts from $k/k_w = 13.74$ and ends at $k/k_w = 14.35$. For the group II orbits with $\Omega_0/k_w c = 2.65$, the wiggler induced velocity $v_w$ is large enough to cause additional couplings between positive and negative-energy space-charge waves and between the $L_e$ and $Sc_+$ modes. The coupling, in Fig. 8(c), between the negative and positive-energy space-charge waves starts from $k/k_w \approx 0.41$ and ends at $k/k_w \approx 24.89$ that have been shown with solid lines. Coupling between the $R_c$ and the $R_e$ modes starts from $k/k_w = 16.19$ and ends at $k/k_w = 17.96$ with $\text{Im}(\omega/k_w c)_{\max} = 0.0344$ at $k/k_w = 17.08$ that have been shown with Circle line. Coupling between the $R_c$ and the $Sc_-$ modes starts from $k/k_w = 28.63$ and ends around $k/k_w \approx 79$ that have been shown with a dotted line. Coupling between the $L_c$ and the $Sc_+$ modes starts from $k/k_w = 32.06$ and ends at $k/k_w = 32.54$ with $\text{Im}(\omega/k_w c)_{\max} = 0.093$ at $k/k_w = 32.31$. This coupling is shown with triangle line in Fig. 8 (c).

## 6. CONCLUSION

In this analysis equilibrium steady-state orbits are found for a electromagnetic wave wiggler FEL with axial magnetic field. The DR for EWW is derived in order to investigate the steady-state orbits. Steady-state orbits show that there is a minimum value of the normalized cyclotron frequency $\Omega_0/\omega_w$ for group I while, there is maximum value of $\Omega_0/\omega_w$ for group II, when angular frequency $\omega_w$ is a constant. The investigation of the effect of the variation of equilibrium orbits shows that in collective or Raman regime there exist an upper limit for the electron beam density when $\omega_w$ is a constant. Our numerical study shows that $\Omega_w < 0$ is also acceptable, this does not indicate a change in the direction of the wiggler field, rather, it indicates a phase shift of $180°$ in EWW. In this case, $v_w$ is in the opposite direction of the previous state. Variation of $k_w$ versus $\gamma_0$ with $\omega_w =$ constant shows that there exist a maximum value of $\gamma_0$ for group II orbits. In the continuation, a general DR (10) is derived for a FEL with a EWW and an axial magnetic field, with all of the relativistic terms related to the transverse velocity included. This DR is solved numerically to study many unstable couplings among all possible modes. In order to analyze the couplings, we introduce the role of the transverse velocity $v_w$ and find that when $v_w$ is zero each wave propagate along the axial magnetic field without interaction with other waves. In the presence of wiggler, transverse velocity causes interaction between modes. Increasing transverse velocity causes the appearance of new couplings between modes. It should be noted that DR (10) is to sixth order in transverse velocity. In DR (10), higher order terms in $v_w$ are from $d\gamma/dt$ and the expansion of the relativistic factor about its equilibrium value $\gamma_0$.

In the group I orbits, the transverse velocity $v_w$ is low, therefore, on the well know FEL coupling between the right wave and the negative-energy space-charge wave is possible. Also, in the group II with strong axial magnetic field only the FEL coupling occurs. For lower values of the normalized cyclotron frequency, in group II orbits, $v_w$ is large and the relativistic terms play important roles, therefore, there exists the FEL coupling with additional coupling between the negative and positive-energy space charge waves and between the $L_e$ and $Sc_+$ modes.

**References:**

[1] H.P.Freund , J.M.Antonsen, "Principles of Free-Electron Lasers,"Chapman and Hall London,1992.

[2] H.P. Freund, R.A. Kehs, V.L. Granatstein "Electron orbits in combined electromagnetic wiggler and axial guide magnetic fields,"IEEE J Quantum Electron, vol.21,pp. 1080 -1082,1985.

[3] H.R.Hiddleston, S.B.Segall , "Equations of motion for a free-electron laser with an electromagnetic pump field and an axial electrostatic field,".EEE J. Quantum Electron,vol.17,pp.1488 -1495,1981.

[4] H.P.Freund, R.A.Kehs ,V.L. Granatstein Linear gain of a free-electron laser with an electromagnetic wiggler and an axial-guide magnetic field. phys. Rev. A,vol. 34, pp.2007-2012,1986.

[5] T.Kwan,J.M. Dawson, "Investigation of the free electron laser with a guide magnetic field," Phys. Fluids, vol 22,pp.1089-1103,1979.

[6] H.P.Freund, P.Sprangle, D.Dillenburg , E.H.da Jornada, R.Schneider ,B.Sand Liberman, " Collective effects on the operation of free-electron lasers with an axial guide field,"Phys. Rev. A ,vol.26, pp.2004-2015,1982.

[7] J.E.Willett, B. Bolon , U-H .Hwang, Y. Aktas " Re-examination of the one-dimensional theory of a Raman free-electron laser," J. Plasma Phys,vol.66,pp.301-313,2001.

[8] H. P. Freund, P.Sprangle "Unstable electrostatic beam modes in free-electron-laser systems," Phys. Rev. A, vol.28, pp.1835-1837,1983.

[9] S.Zhang, J.Elgin , "Propagation of the equilibrium electron beam in a free-electron laser with an axial guide magnetic field," Phys. Rev. E.vol.55, pp. 4684, 1997.

[10] J.T.Donohue, J.L.Rullier, "Electron trajectories in a helical free-electron laser with or without an axial guide field, " Nucl. Instrum. Methods Phys. Res A, vol, 507, pp.56-60, 2003.

[11] K.D.Misra, P.K.Mishra ," Generalized theory of a free-electron laser in a helical wiggler and guide magnetic fields using the kinetic approach,"Phys Plasmas ,vol.9, pp.330-339,2002.

[12] P. K. Mishra," Kinetic Description of a Low-Gain Free-Electron Laser using the Einstein Coefficient," Method. Laser Physics, vol.15, pp.679 -685,2005.

[13] T.Mohsenpour, B.Maraghechi , S.Mirzanejhad, "Unstable coupled-mode structures in a one-dimensional Raman free-electron laser," Phys. Plasmas, vol.14, pp.053106 -121, 2007.

[14] H.Mahdavi, B. Maraghechi, T.Mohsenpour , "Self-field effects on the wave-mode couplings in a free-electron laser. Plasma Phys, "Control. Fusion, vol.51, pp.075007-0750021,2009.

[15] P.Sprangle, V.L.Granatstein ," Stimulated cyclotron resonance scattering and production of powerful submillimeter radiation'" Appl. Phys. Lett,"vol.25,pp.377-379,1974.

[16] A.T.Lin, J.M.Dawson, "Nonlinear saturation and thermal effects on the free electron laser using an electromagnetic pump,"Phys. Fluids,vol. 23, pp.1224-1228,1984

[17] J.Gea-Banacloche, G.T.Moore, R.R.Schlicher, M.O Scully, H.Walther, " Soft X-ray free-electron laser with a laser undulator'"IEEE J. Quantum Electron, vol.23, pp.1558-1570,1987.

[18] B.G.Danly, G.Bekefi, R.C.Davidson, R.J.Temkin, T.M.Tran, J.S.Wurtele, "Principles of gyrotron powered electromagnetic wigglers for free-electron lasers, " IEEE J. Quantum Electron,vol. 23,pp. 103-116,1987.

[19] Y.Carmel, V.L.Grantstein, A.Gover, " Demonstration of a Two-Stage Backward-Wave-Oscillator Free-Electron Laser'" Phys. Rev. Lett,vol.51,pp.566 -569,1983.

[20] P.Sprangle, B.Hafizi, J.R.Penano ," Laser-pumped coherent x-ray free-electron laser" Phys. Rev. ST Accel. Beams, vol. 12, pp.050702-12,2009.

[21] A.,Bacci, C.Maroli, V.Petrillo, A.R.Rossi,L. Serafini, P.Tomassini ," Compact X-ray free-electron laser based on an optical undulator," Nucl. Instrum. Methods Phys Res. A ,vol.587, pp.388 –397,2008.

[22] A.Bacci, M.Ferraria, C.Maroli, V.Petrillo, L.Serafini, "Transverse effects in the production of x rays with a free-electron laser based on an optical undulator,"Phys. Rev. ST Accel. Beams, vol.9, pp.060704 -0607049,2006.

[23] Y.Seo, " The nonlinear evolution of a free electron laser with electromagnetic wigglers, "Phys. Fluids B,vol. 3,pp. 797-.810,1991.

[24] M.Olumi, B.Maraghechi, M.H.Rouhani ," Traveling waves in a free-electron laser with an electromagnetic wiggler. Plasma Phys. Control. Fusion, vol. 53, pp.015010-21,2010.

[25] B. Maraghechi and N. Sepehri Javan, "Raman free-electron laser with longitudinal electrostatic wiggler and annular electron beam" Phys. Plasmas vol. 8, pp.4193, 2001.

[26] I. N. Kartashov, M. V. Kuzelev, and N. Sepehri Javan, "Unsteady processes during stimulated emission from a relativistic electron beam in a quasi-longitudinal electrostatic pump field" Plasma Phys. Rep. vol. 31, pp.244, 2005.

[27] N. Sepehri Javan, "Free electron laser with bunched relativistic electron beam and electrostatic longitudinal wiggler" Phys. Plasmas vol. 17, pp.063105, 2010.

[28] N. Sepehri Javan, "Lasing conditions of a free electron laser with helical wiggler" Phys. Plasmas vol.16, pp.123109 2009.

[29] I. N. Kartashov, M. V. Kuzelev, and A. A. Rukhadze, "Single-particle Cherenkov effect in a bounded spatial region" Tech. Phys. 51, 151(2006).

[30] I. N. Kartashov, M. V. Kuzelev, A. A. Rukhadze, and N. Sepehri Javan, "Unsteady processes during stimulated emission from a relativistic electron beam in a quasi-longitudinal electrostatic pump field" Tech. Phys. Vol.50, pp.298 2005.

[31] N. Sepehri Javan, "Threshold conditions for lasing of a free electron laser oscillator with longitudinal electrostatic wiggler " Phys. Plasmas vol.19, pp.122306 (2012).

[32] A.Taghavi, M. Esmaeilzadeh and M. S. Fallah, "Chaotic electron trajectories in an electromagnetic wiggler free-electron laser with ion-channel guiding " Phys. Plasmas vol.17, pp.093103 (2010).

[33] N. Nasr, H. Mehdian and A. Hasanbeigi , "Controlling chaotic behavior of the equilibrium electrons by simultaneous using of two guiding fields in a free-electron laser with an electromagnetic-wave wiggler " Phys. Plasmas vol.18, pp.043104 (2011).

[34] M. Esmaeilzadeh, V. Ghafouri, A. Taghavi, and E. Namvar, "Electron trajectories and gain for an electromagnetic wiggler with ion-channel guiding" Phys. Plasmas vol.13, pp.043103 (2006).

[35] H. Mehdian, A. Hasanbeigi, and S. Jafari, "Free-electron laser harmonic generation in an electromagnetic-wave wiggler and ion channel guiding " Phys. Plasmas vol.17, pp.023112 (2010).

[36] H. E. Amri and T. Mohsenpour, "Effects of electromagnetic wiggler and ion channel guiding on equilibrium orbits and waves propagation in a free electron laser" Phys. Plasmas vol.23, pp.022101 (2016).

[37] H. Mehdian and S. Jafari, "A comparison between electron orbits for both an axial magnetic field and an ion-channel guiding in a FEL with an electromagnetic wave wiggler" J. Plasma physics vol.74 pp.178 (2008).

[38] L.Shenggang , R.J.Barker, G.Hong, Y.Yang ," Electromagnetic wave pumped ion-channel free electron laser," Nucl. Instrum. Methods Phys. Res. A,vol. 475,pp.153-157,2001.

[39] Larsson and V. Thomée, Eds., "Introduction," in *Partial Differential Equations with Numerical Methods*, Berlin, Heidelberg: Springer, 2003, pp. 1–14. doi: 10.1007/978-3-540-88706-5_1.

[40] C. M. Elliott and T. Ranner, "Finite element analysis for a coupled bulk–surface partial differential equation," *IMA J. Numer. Anal.*, vol. 33, no. 2, pp. 377–402, Apr. 2013, doi: 10.1093/imanum/drs022.

[41] J. H. Coggon, "Electromagnetic and electrical modeling by the finite element method," *GEOPHYSICS*, vol. 36, no. 1, pp. 132–155, Feb. 1971, doi: 10.1190/1.1440151.

[42] E. L. Wilson and R. E. Nickell, "Application of the finite element method to heat conduction analysis," *Nucl. Eng. Des.*, vol. 4, no. 3, pp. 276–286, Oct. 1966, doi: 10.1016/0029-5493(66)90051-3.

[43] H. Rahmaninejad, T. Pace, S. Bhatt, B. Sun, and P. Kekenes-Huskey, "Co-localization and confinement of ecto-nucleotidases modulate extracellular adenosine nucleotide distributions," *PLOS Comput. Biol.*, vol. 16, no. 6, pp. 1–31, Jun. 2020, doi: 10.1371/journal.pcbi.1007903.

[44] F. Assous, P. Degond, E. Heintze, P. A. Raviart, and J. Segre, "On a Finite-Element Method for Solving the Three-Dimensional Maxwell Equations," *J. Comput. Phys.*, vol. 109, no. 2, pp. 222–237, Dec. 1993, doi: 10.1006/jcph.1993.1214.

[45] S. S. Rao, *The Finite Element Method in Engineering*. Butterworth-Heinemann, 2017.

[46] J. Pomplun, S. Burger, L. Zschiedrich, and F. Schmidt, "Adaptive finite element method for simulation of optical nano structures," *Phys. Status Solidi B*, vol. 244, no. 10, pp. 3419–3434, 2007, doi: 10.1002/pssb.200743192.

[47] T. Pace, H. Rahmaninejad, B. Sun, and P. M. Kekenes-Huskey, "Homogenization of Continuum-Scale Transport Properties from Molecular Dynamics Simulations: An Application to Aqueous-Phase Methane Diffusion in Silicate Channels," *J. Phys. Chem. B*, vol. 125, no. 41, pp. 11520–11533, Oct. 2021, doi: 10.1021/acs.jpcb.1c07062.

[48] P. R. F. Teixeira and A. M. Awruch, "Numerical simulation of fluid–structure interaction using the finite element method," *Comput. Fluids*, vol. 34, no. 2, pp. 249–273, Feb. 2005, doi: 10.1016/j.compfluid.2004.03.006.

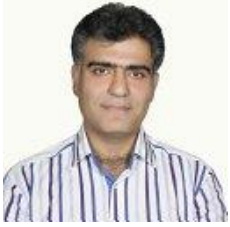

**Hassan Ehsani Amri** :Recived his PhD from the science and research branch of Islamic Azad University in Fabruary 2011.He was group leader of physics and photonic department of Islamic Azad University of Nour branch from 2011-2016.Now he is with the physics department of Islamic Azad university as an assistant professor and his research area are laser spectroscopy, photonics and free electron laser.

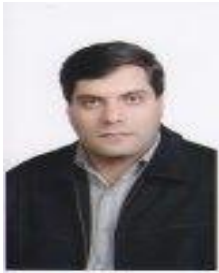

**Taghi Mohsenpour:** Recived his PhD degree from the Amirkabir University of Technology in February 2010.He is now with the physics department of University of Mazandaran as an associated professor and group leader .His research area are free electron laser and quantum plasmas.